\documentclass{article}
\usepackage{gese}

\usepackage{times}
\usepackage{soul}
\usepackage{url}
\usepackage[hidelinks]{hyperref}
\usepackage[utf8]{inputenc}
\usepackage[small]{caption}
\usepackage{graphicx}
\usepackage{amsmath}
\usepackage{amsthm}
\usepackage{booktabs}
\usepackage{algorithm}
\usepackage{algorithmic}
\usepackage{multirow} 
\usepackage{amsmath}
\usepackage{amsfonts}

\usepackage{amsmath}
\DeclareMathOperator*{\argmax}{arg\,max}

\title{Generate to Explore, Select to Exploit: Aligning LLM-based Headline Generation with Personalized Recommendation}

\author{
Yi Chen
\and
Rufeng Cheng
\and
Qiang Xie$^{*}$
\and
Tao Li$^{*}$
\affiliations
Baidu Inc., Beijing, China
\emails
\{chenyi32, chengrufeng, xieqiang, litao\}@baidu.com
\emails
{$^{*}$Corresponding author}
}

\begin{document}

\maketitle

\begin{abstract}


In industrial recommendation feeds, presenting a static headline for an item often fails to satisfy the diverse, multimodal interests of the user population, particularly suppressing the needs of long-tail audiences. While Large Language Models (LLMs) have been integrated into recommendation for content understanding or ranking, directly optimizing them to output a single best headline typically leads to mode collapse---converging to generic patterns that satisfy average tastes but miss specific latent intents. To bridge this gap, we introduce GESE (\textbf{G}enerate to \textbf{E}xplore, \textbf{S}elect to \textbf{E}xploit), a framework operating at the system's presentation layer that decouples personalization into generative exploration and selective exploitation. First, we treat the LLM as a probabilistic explorer, utilizing Group Sequence Policy Optimization (GSPO) with a hierarchical reward mechanism to generate a candidate set that maximizes the semantic coverage of potential user interests. Subsequently, a lightweight, real-time feedback-aware selector acts as the exploiter, identifying the optimal realization from the candidate pool based on instant contextual signals. Extensive deployment on a commercial platform with over 100 million daily active users demonstrates that GESE significantly outperforms state-of-the-art baselines, achieving a 2.57\% lift in CTR and 0.87\% in dwell time. These results validate that decoupling diversity-oriented generation from precision-oriented selection offers a robust blueprint for aligning generative AI with dynamic user utility.

\end{abstract}

\section{Introduction}

\begin{figure}[ht]
    \centering
    \includegraphics[width=0.95\columnwidth]{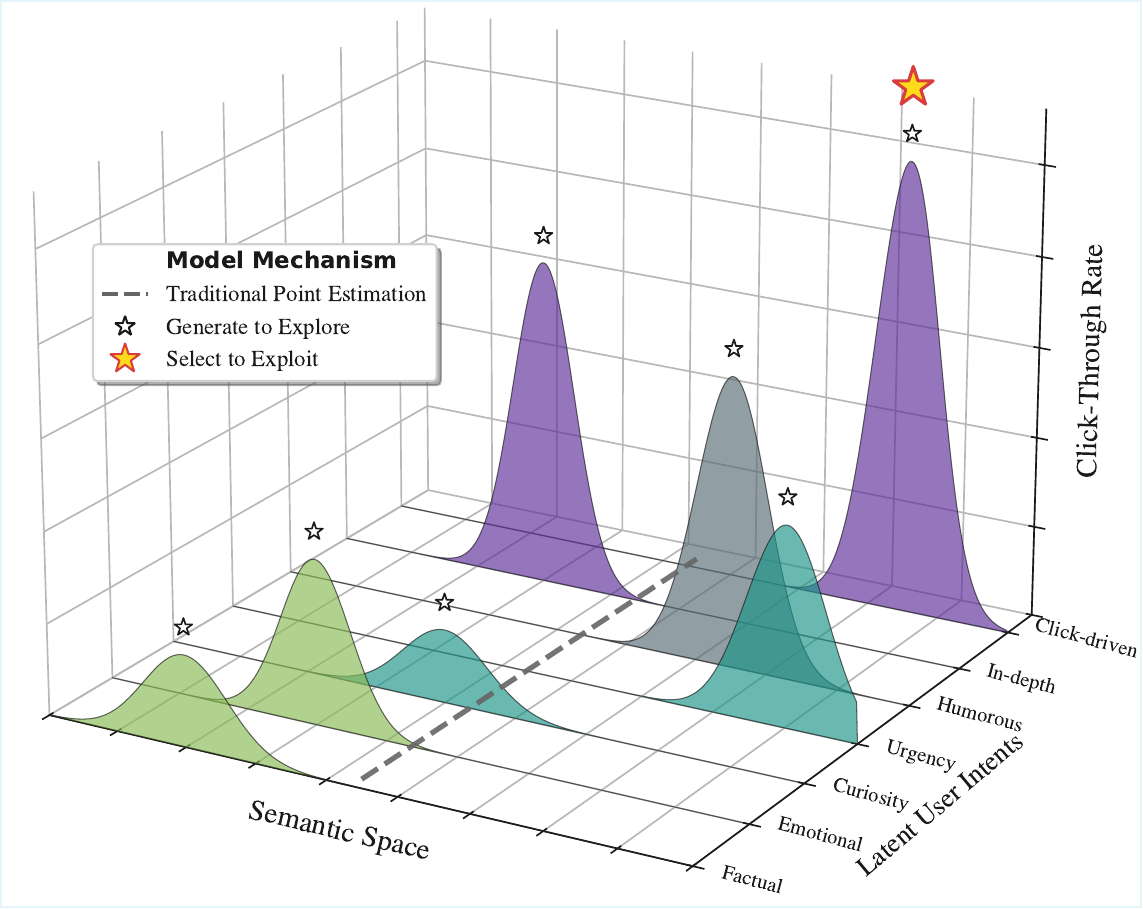}
    \caption{Static Point Estimation vs. Distributional Coverage. Traditional methods (gray line) suffer mode collapse on the multimodal interest landscape. In contrast, GESE decouples the task: Exploration (small stars) maximizes semantic coverage across diverse peaks, while Exploitation (gold star) leverages real-time signals to pinpoint the optimal realization.}
    \label{fig:teaser}
\end{figure}

In the architecture of modern industrial recommendation systems, the integration of Large Language Models (LLMs) has predominantly focused on upstream tasks, such as enhancing item representations via embeddings or modeling sequential user behaviors for retrieval. However, a pivotal problem remains underexplored: the Presentation Layer. While a ranking model may successfully retrieve a relevant item, the static presentation—specifically the headline displayed—often fails to resonate with the heterogeneous nature of the user base. A single editorial headline typically cannot satisfy the diverse, multimodal interests of millions of users, particularly suppressing the needs of long-tail audiences.

The industry standard frames personalized generation as a precise point-estimation problem, training LLMs to output a single optimal headline for a given user profile. However, this pursuit of precision rests on a fragile assumption: that user intent is unimodal and noise-free. In reality, user interests in open-domain feeds are inherently multimodal and dynamic. As visualized in Figure \ref{fig:teaser}, a user's latent preference for an item often splits into distinct intent modes (e.g., Factual vs. Emotional). Under the aleatoric uncertainty of sparse industrial data\cite{10.1145/3626772.3657684,jiang-etal-2025-reclm}, standard alignment methods (like SFT or DPO) function as a static point estimator (represented by the dashed gray line). They tend to regress to the mean of the distribution, leading to mode collapse—generating generic, safe patterns that cut through the semantic space but miss the specific high-utility peaks of individual intents. This results in the Echo Chamber effect, where long-term engagement is degraded by repetitive, average-quality content.

Current approaches struggle to resolve this tension because they couple generation and ranking into a single black-box optimization process. Methods based on Supervised Fine-Tuning (SFT) tend to mimic the mode of the training distribution, lacking the incentive to explore the semantic space\cite{Li2024PreservingDI}. Meanwhile, standard Reinforcement Learning from Human Feedback (RLHF) approaches, such as DPO\cite{10.5555/3666122.3668460}, often optimize for a scalar reward that inadvertently encourages homogeneity. By collapsing complex user preferences into a single point-wise score, these methods sacrifice the semantic richness necessary to capture latent user interests\cite{NEURIPS2024_f25d75fc}. Consequently, end-to-end models have no mechanism to hedge their bets against profile noise.

In this paper, we propose to decouple the recommendation problem into two distinct phases: Generation as Exploration and Selection as Exploitation. We introduce GESE (Generate to Explore, Select to Exploit), a framework designed for the fusion stage of recommendation pipelines. 

Instead of demanding the LLM to perform the final ranking, we position the LLM as a generator of a diverse hypothesis space. We employ Group Sequence Policy Optimization (GSPO)\cite{zheng2025groupsequencepolicyoptimization} to train the LLM, shifting its objective from minimizing token perplexity to maximizing the distributional coverage of potential user interests (the white stars in Figure \ref{fig:teaser}).
Subsequently, a real-time feedback-aware selector acts as the exploiter. Unlike ID-based methods that rely on static history, this lightweight module utilizes instant contextual signals—such as immediate session feedback and real-time user meta-tags—to identify the optimal realization (the gold star) from the diverse pool.

This decoupled paradigm offers a theoretically grounded solution to the Personalization Paradox: by generating for the group (diversity) and selecting for the individual (precision), we utilize generative diversity as a mathematical hedge against profile uncertainty. Our contributions are as follows:

\begin{itemize}

    \item \textbf{Framework Decoupling}: We propose the GESE framework, which redefines personalized headline generation by separating the responsibility of semantic exploration (LLM) from preference exploitation (Ranking Model), specifically addressing the robustness issue in the presentation layer of noisy industrial environments.
    \item \textbf{Reward Function Optimization via GSPO}: We introduce a novel alignment strategy utilizing a hierarchical reward mechanism that synthesizes generation diversity, predicted CTR (Click-Through Rate), and content faithfulness. This approach not only prevents mode collapse by ensuring the candidate set covers the diverse interest spectrum of the target audience, but also significantly enhances computational efficiency and reduces inference costs by guiding the model to generate the entire candidate set in a single inference pass.
    \item \textbf{Industrial Verification}: We deploy the proposed method on a commercial recommendation platform with over 100 million DAUs. Extensive offline evaluations and online A/B testing demonstrate that our approach significantly outperforms strong baselines (including standard SFT and DPO), achieving a 2.57\% lift in CTR and 0.87\% in dwell time, validating that diversity is indeed a prerequisite for superior personalization.
    
\end{itemize}

\section{Related Work}

\subsection{LLM Alignment in Recommendation}

The integration of LLMs into recommender systems has shifted the paradigm from ID-based matching to generative content understanding\cite{10506571,geng2023recommendationlanguageprocessingrlp}. Early approaches primarily utilized Supervised Fine-Tuning (SFT) to adapt LLMs for tasks like rating prediction or explanation generation\cite{o'mahony2024attributing}. While SFT effectively instills domain knowledge, it suffers from the fundamental limitation of behavior cloning: the model learns to mimic the average distribution of the training data, often leading to generic, safe, and undistinguished outputs—a phenomenon known as the regression to the mode\cite{o'mahony2024attributing}.


To transcend simple imitation, recent works have adopted Reinforcement Learning from Human Feedback (RLHF), specifically Direct Preference Optimization (DPO) and its variants\cite{10.5555/3666122.3668460,pmlr-v235-ethayarajh24a}. These methods align LLMs with user preferences by optimizing a contrastive loss between chosen and rejected pairs. However, a critical disconnect remains in industrial applications: standard alignment methods treat user preference as a static, point-wise label. They optimize for the probability of a single best response, assuming a deterministic mapping from user profile to user intent\cite{zhou-etal-2024-beyond}. As highlighted by\cite{Wang_2025}, this point-wise maximization is brittle against profile noise; when the user representation is ambiguous, DPO-trained models tend to overcommit to a specific (often incorrect) interest cluster or collapse into low-diversity patterns to maximize the scalar reward. In contrast, our work posits that alignment in recommendation must be distributional rather than point-wise. By employing Group Sequence Policy Optimization (GSPO)\cite{zheng2025groupsequencepolicyoptimization}, we move beyond optimizing the expected value of a single item to optimizing the coverage of the user's potential interest spectrum.


\subsection{Diverse Generation}

Diversity in recommendation has traditionally been treated as a post-hoc constraint, enforced via re-ranking algorithms like Maximal Marginal Relevance (MMR) or Determinantal Point Processes (DPP)\cite{articleATI,10.5555/3327345.3327465}. These methods operate on a fixed candidate pool retrieved by upstream models. However, in the era of Generative Recommendation, diversity must be intrinsic to the generation process itself. If the LLM collapses to a single mode during generation, no amount of post-hoc re-ranking can recover the lost semantic nuances\cite{zhang2025moslimaligndiversepreferencesprompts}.


Recent efforts in diversity-aware generation have explored entropy-regularized decoding or prompt-based diversification strategies\cite{ye2024scalarrewardmodellearning,10.5555/3692070.3694392}. While these methods increase lexical variety, they often sacrifice semantic faithfulness and relevance, leading to the diversity-quality trade-off\cite{zhou-etal-2025-balancing}. More importantly, existing generative frameworks typically couple the generation and ranking objectives end-to-end. They demand the generator to be simultaneously diverse (to explore) and precise (to exploit), objectives that are often mathematically conflicting within a single policy network\cite{10.5555/3737916.3741662}. Our GESE framework addresses this by structurally decoupling these responsibilities: the LLM is released from the burden of ranking to focus on exploration via diverse hypothesis generation, while a dedicated real-time feedback model handles the exploitation via context-aware selection. This separation allows us to achieve robustness against profile noise without compromising the precision required for industrial CTR optimization.


\section{Proposed Method}

\begin{figure*}[ht]
\centering
\includegraphics[width=0.98\textwidth]{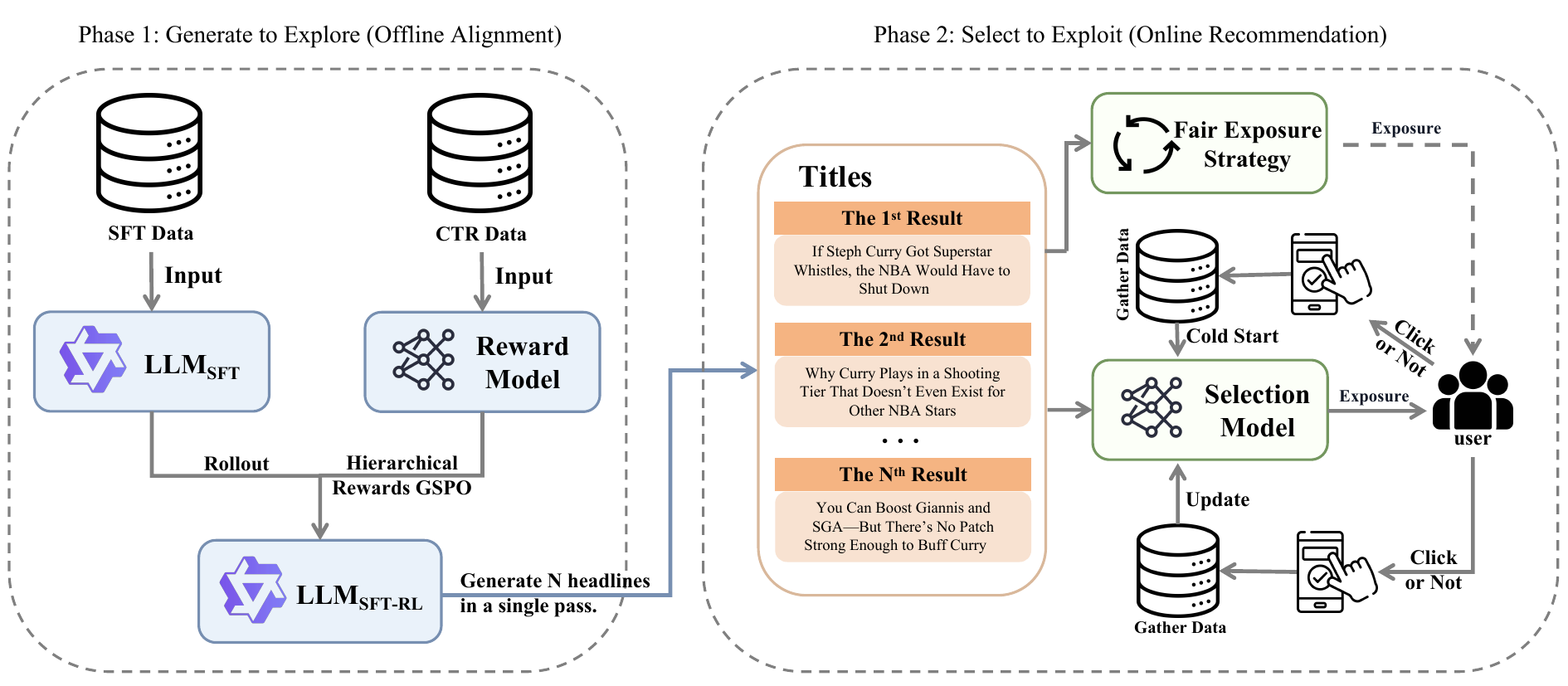}
\caption{The GESE Framework: Decoupling Exploration from Exploitation for Robust Personalized Headline Generation. 
Phase 1 (Offline) trains an LLM to generate a diverse set of candidate headlines via Group Sequence Policy Optimization (GSPO), using SFT data and CTR signals to shape a hierarchical reward that values diversity, faithfulness, and engagement potential. Phase 2 (Online) deploys a lightweight, real-time feedback Selection Model to pick the optimal headline from the candidate pool based on real-time user feedback. A Fair Exposure Strategy ensures that the online model receives sufficient user impression–click data as cold-start training signals, allowing the online selection module to effectively fit user-level click behaviors.}
\label{fig:framework}
\end{figure*}

Formally, standard personalized generation is a point-wise estimation task: training a model $\pi_\theta(y|\mathbf{x}, c)$ to predict a single optimal headline $y$ given a user profile $\mathbf{x}$. This approach fails when $\mathbf{x}$ serves as a noisy, sparse proxy for the user's true, multimodal latent intent $z \sim P(z|\mathbf{x})$, inevitably causing the model to suffer from mode collapse.

To address this, we reformulate the problem as set-wise optimization via our GESE (Generate to Explore, Select to Exploit) framework. We decouple the task into two phases. Generative exploration is optimized for distributional coverage. Its goal is to produce a candidate set $\mathcal{Y}=\{y_1, \dots, y_K\}$ that is highly likely to contain an optimal match for any potential user intent $z$. This transforms the objective to maximizing the expected utility of the best candidate within the set, where $u(y, z)$ is the utility function:

    \begin{equation}
        \max_\theta \mathbb{E}_{z \sim P(z|\mathbf{x})} \left[ \max_{y_k \in \mathcal{Y}} u(y_k, z) \right]
    \end{equation}
    
 Selective exploitation, A separate real-time-aware ranker $\mathcal{S}_\phi$, leverages real-time contextual signals to identify and serve the optimal realization $\hat{y}$ from the generated pool: 

\begin{equation}
 \hat{y} = \argmax_{y \in \mathcal{Y}} \mathcal{S}_\phi(y, \mathbf{x}, c).
\end{equation}

This framework relieves the LLM of noise-sensitive ranking duties, allowing it to focus on semantic exploration while the ranker ensures precision.

\subsection{Generative Exploration}

To align the Large Language Model (LLM) with the objective of constructing a high-coverage candidate set $\mathcal{Y}$, we employ a three-stage training pipeline. This pipeline progressively transitions the model from adhering to linguistic instructions to optimizing for a complex, multi-objective utility function.

\paragraph{Supervised Instruction Tuning (SFT).}

The foundation of our explorer is a Qwen3-14B backbone. While the pre-trained model possesses general linguistic knowledge, it lacks two critical capabilities required for our industrial pipeline: (1) adhering to strict output formatting (e.g., JSON), and (2) Single-pass Multi-candidate Generation. 

Existing methods often rely on repeated sampling (calling the LLM $K$ times) to obtain multiple candidates. This approach is not only computationally prohibitive for real-time serving but also suboptimal for diversity, as independent samples tend to cluster around the mode of the distribution.

To address this, we construct a dataset $\mathcal{D}_{SFT}$ comprising 30k samples, where the target $y^*$ is a structured list of distinct headlines. We fine-tune the model to generate the entire candidate set $\mathcal{Y} = \{y_1, \dots, y_K\}$ in a single inference pass. The optimization objective is:

\begin{equation}
    \mathcal{L}_{SFT}(\theta) = - \mathbb{E}_{(c, \mathbf{x}, \mathcal{Y}) \sim \mathcal{D}_{SFT}} \sum_{t=1}^{|\mathcal{Y}|} \log \pi_\theta(\mathcal{Y}_t | c, \mathbf{x}, \mathcal{Y}_{<t})
\end{equation}

This warm start stage ensures the policy $\pi_{SFT}$ learns to output diverse options within a single context window, significantly reducing inference costs while establishing the structural basis for subsequent exploration.

\paragraph{Latent Interest Calibration via Reward Modeling.}
\label{sec:generative_exploration}

To guide the exploration process, we require a robust proxy for user preference. We train a Click-Through Rate (CTR) Reward Model (RM) that predicts the scalar utility of a generated headline.

Unlike standard RMs that rely solely on generic semantic embeddings, we design a Hybrid-Feature Reward Architecture. To ensure semantic alignment with our generator, we utilize the Qwen-Embedding model as the backbone encoder $E_\phi(\cdot)$. Specifically, we feed the concatenated context and headline sequence $c \oplus y$ into the encoder and extract the last token's hidden state as the dense semantic representation, preserving the full contextual understanding of the Qwen architecture.
Simultaneously, we explicitly model display-bias features $\mathbf{f}_{disp}$ (e.g., character length, punctuation density) which significantly impact industrial CTR.

Let $\mathbf{h}_{sem} \in \mathbb{R}^d$ denote the extracted semantic vector and $\psi(\cdot)$ be a feature projection network. The predicted reward $r_\phi(y, c)$ is formulated as a non-linear fusion of semantic and display signals:

\begin{equation}
    r_\phi(y, c) = \text{MLP}_{head} \left( \underbrace{E_\phi(c \oplus y)[\texttt{EOS}]}_{\mathbf{h}_{sem}} \parallel \underbrace{\psi(\mathbf{f}_{disp})}_{\mathbf{h}_{feat}} \right)
\end{equation}
    
where $\parallel$ denotes vector concatenation, and $[\texttt{EOS}]$ signifies the extraction of the representation at the end-of-sequence token position. We utilize 8 million pairs of high-confidence interaction logs $\mathcal{D}_{RM} = \{(y_w, y_l)\}$, where $y_w$ (chosen) has a significantly higher CTR than $y_l$ (rejected). The model is optimized via the Bradley-Terry pairwise ranking loss:

\begin{equation}
    \mathcal{L}_{RM}(\phi) = - \mathbb{E}_{(y_w, y_l) \sim \mathcal{D}_{RM}} \left[ \log \sigma (r_\phi(y_w) - r_\phi(y_l)) \right]
\end{equation}

This RM serves as the primary signal for Attractiveness in our hierarchical reward function.

\paragraph{Exploration via GSPO.}

Standard Reinforcement Learning methods like PPO often suffer from high variance and require a separate Value Network, while DPO is limited to pairwise comparisons, restricting the model's ability to explore the global semantic space. To address the Echo Chamber effect and enforce diversity, we adopt Group Sequence Policy Optimization (GSPO).

GSPO operates by sampling a group of $G$ outputs $\{y_1, \dots, y_G\}$ for each prompt $(c, \mathbf{x})$ from the current policy $\pi_\theta$. Instead of relying on an absolute scalar reward, GSPO optimizes the policy based on the \textit{relative advantage} of each candidate within its group.

Formally, for each sampled output $y_i$ in the group, we compute a total reward $R(y_i)$ (synthesizing CTR, diversity, and faithfulness). The advantage $A_i$ is derived by normalizing the rewards within the group:

\begin{equation}
    A_i = \frac{R(y_i) - \mu(\{R(y_j)\}_{j=1}^G)}{\sigma(\{R(y_j)\}_{j=1}^G) + \epsilon}
\end{equation}

where $\mu$ and $\sigma$ are the mean and standard deviation of rewards within the group. The policy $\pi_\theta$ is then updated to maximize the following objective:

\begin{equation}
\label{eq:gspo}
\mathcal{L}_{\text{GSPO}}(\theta) = \mathbb{E}_{q, \{y_i\}} \left[ 
\frac{1}{G} \sum_{i=1}^G \left( \mathcal{L}_i^{\text{clip}} - \beta D_{\text{KL}}[\pi_\theta \| \pi_{\text{ref}}] \right) 
\right]
\end{equation}

where $\mathcal{L}_i^{\text{clip}} = \min \left( \rho_i A_i, \text{clip}(\rho_i, 1-\epsilon, 1+\epsilon) A_i \right)$, $\rho_i = \frac{\pi_\theta(y_i|c)}{\pi_{\theta_{old}}(y_i|c)}$ is the importance sampling ratio. The rationale for employing GSPO lies in its intrinsic mechanism for diversity promotion. By utilizing group-wise normalization, GSPO creates a local competition environment within the generated batch. For the model to minimize the loss, it cannot simply generate $G$ identical safe headlines, as this would result in a zero standard deviation and undefined advantages. Instead, the objective function mathematically compels the policy to differentiate the candidates, pushing the model to explore the boundaries of the semantic space to find distinct, high-reward strategies.

\subsection{Hierarchical Reward Shaping}

Since our policy $\pi_\theta$ generates a candidate set $\mathcal{Y} = \{y_1, \dots, y_K\}$ in a single pass, we define a hierarchical reward $R(\mathcal{Y}, c, \mathbf{x})$ that synthesizes individual semantic utility with collective distributional diversity.

\paragraph{Semantic Utility.}

For each headline $y_k$, we compute a utility score $U(y_k)$ balancing three objectives: 1) Attractiveness $r_{ctr}$: The predicted CTR from our Reward Model, clipped for stability. 2) Faithfulness $r_{qual}$: A hard constraint using a discrimination model $\mathcal{M}_{qual}$, returning $1.0$ if $y_k$ is factually consistent and safe, else a penalty $\eta_{penalty}$. 3) Novelty $r_{nov}$: To discourage simply copying the editorial title $y_{org}$, we utilize the Normalized Edit Distance, defined as: 

\begin{equation}
    r_{nov} = \text{Levenshtein}(y_k, y_{org}) / \max(|y_k|, |y_{org}|).
\end{equation}

The aggregated utility is a weighted sum:

\begin{equation}
    U(y_k) = r_{ctr}(y_k) + \lambda_1 r_{len}(y_k) + \lambda_2 r_{qual}(y_k) + \lambda_3 r_{nov}(y_k)
\end{equation}

\paragraph{Set-wise Diversity.}
To prevent mode collapse—where the model generates $K$ identical high-CTR captions—we explicitly penalize redundancy. We define set-wise redundancy using the Intersection-over-Union (IoU) of character sets $S(y_k)$, the diversity bonus $R_{div}$ is scaled inversely to this redundancy:

\begin{equation}
     \text{Red}(\mathcal{Y}) = \frac{|\bigcap_{k} S(y_k)|}{|\bigcup_{k} S(y_k)| + \epsilon}. 
\end{equation}

\paragraph{Final Objective.}
The final reward $R_{total}$ for GSPO optimization is the Z-score normalized synthesis of average precision and set-wise diversity:

\begin{equation}
    R_{total}(\mathcal{Y}) = \underbrace{\frac{1}{K} \sum_{k=1}^K U(y_k)}_{\text{Average Utility}} + \alpha \cdot \max\left(0, (1 - \text{Red}(\mathcal{Y})) \cdot \beta \right)
\end{equation}

This objective compels the policy to explore novel, high-CTR angles (Novelty + Driver) while ensuring candidates remain diverse (Expander) and faithful (Guardrail).

\subsection{Online Exploitation}
\label{sec:online_exploitation}

While the LLM acts as a semantic explorer providing a diverse candidate set $\mathcal{Y}$, identifying the optimal title for a specific user context requires real-time exploitation. This phase faces the classic Explore-Exploit dilemma: we must accurately estimate the CTR of new candidates (Exploration) while maximizing immediate user engagement (Exploitation). We address this via a two-stage mechanism: a Confidence-Aware Bandit for cold-start and a Deep Neural Ranker for mature serving.

\paragraph{Cold Start.}

New generated headlines lack interaction history. To ensure fair exposure and rapid convergence, we propose a Confidence-Aware Thompson Sampling (CATS) strategy. We model the click probability $\theta_i$ of each candidate $y_i$ (including the original title $y_{org}$) as a Beta distribution: $\theta_i \sim \text{Beta}(\alpha_i, \beta_i)$, where $\alpha_i = 1 + \text{clicks}_i$ and $\beta_i = 1 + \text{views}_i - \text{clicks}_i$.

Standard Thompson Sampling selects $y_k = \argmax_i \tilde{\theta}_i$ where $\tilde{\theta}_i \sim \text{Beta}(\alpha_i, \beta_i)$. However, in industrial settings, unbounded exploration is risky. We impose a Statistical Termination Mechanism to switch from exploration to exploitation dynamically. 

Let $P(\text{win}_i) = P(\theta_i > \theta_{org} | \mathcal{D})$ denote the Bayesian confidence that candidate $i$ outperforms the original title. The active candidate set $\mathcal{S}_t$ available for Thompson Sampling is defined by the following Rotation Criteria:

\begin{equation}
\label{eq:rotation}
\small
\begin{aligned}
    i \in \mathcal{S}_t \iff & \ N_i \leq \tau_{warm} \\
    & \lor \big( P(\text{win}_i) < \delta_{high} \land N_i \leq \tau_{max} \land \bar{\theta}_i > \bar{\theta}_{org} \big) \\
    & \lor \big( P(\text{loss}_i) < \delta_{low} \ \land N_i \leq \tau_{max} \land \bar{\theta}_i < \bar{\theta}_{org} \big)
\end{aligned}
\end{equation}

Where $N_i$ is the impression count and $\bar{\theta}_i$ is the empirical CTR. Here, $\tau_{warm}$ represents the minimum sample size required for statistical validity, $\tau_{max}$ acts as a safety budget to cap the exploration cost, and $\delta_{high} / \delta_{low}$ are high/low confidence thresholds for declaring a candidate as a winner or loser, respectively.

If a candidate $y_i$ satisfies Eq. (\ref{eq:rotation}), its display probability is determined by Thompson Sampling; otherwise, the exploration terminates: the candidate is either permanently discarded (if proven suboptimal) or fixed as the winner (if proven superior). This mechanism ensures that high-potential titles accumulate sufficient data while inferior ones are pruned early.

\paragraph{Online Selection.}

Once candidates accumulate sufficient interaction data via the Bandit phase, we transition to a supervised Context-Aware Ranker for fine-grained exploitation. This transition allows us to move beyond simple ID-based statistics and capture high-order feature interactions between the user's heterogeneous state and the semantic nuances of the generated headlines.

\begin{table}[h]
\centering
\caption{Feature Set for Online Selection Model}
\label{tab:features}
\resizebox{\columnwidth}{!}{%
\begin{tabular}{@{}llp{4.5cm}@{}}
\toprule
\textbf{Domain} & \textbf{Feature Name} & \textbf{Description} \\ \midrule
\multirow{5}{*}{\textbf{User Profile}} & \texttt{uid}, \texttt{cuid} & Unique User/Device Identifiers (Embedding) \\
 & \texttt{gender}, \texttt{age} & Basic Demographic Attributes \\
 & \texttt{education} & User's Educational Background Level \\
 & \texttt{consumption} & Estimated Purchasing Consumption Level \\
 & \texttt{group} & User Group Segment \\ \midrule
\multirow{4}{*}{\textbf{User Interest}} & \texttt{user\_metags\_20} & Sequence of last 20 fine-grained interest tags \\
 & \texttt{user\_metags\_top1} & Most frequent fine-grained interest tag \\
 & \texttt{user\_mecate\_v2} & Sequence of coarse-grained category interests \\
 & \texttt{mecates\_v2\_top1} & Top-1 coarse-grained category preference \\ \midrule
\multirow{2}{*}{\textbf{Candidate}} & \texttt{original\_title} & Semantic Embedding of the editorial title \\
 & \texttt{generated\_title} & Semantic Embedding of the LLM hypothesis \\ \bottomrule
\end{tabular}%
}
\end{table}

We construct the input instance by aggregating features from three views: User Profile, Dynamic Interest, and Candidate Semantics (detailed in Table \ref{tab:features}). Let $\mathbf{x}_{u}$ and $\mathbf{x}_{y}$ denote the sparse feature vectors for user $u$ and candidate headline $y$, respectively.
We first employ an Embedding Layer $\mathbf{E}(\cdot)$ to project sparse categorical features into dense low-dimensional spaces. Continuous features and semantic embeddings are concatenated directly.

The probability of user $u$ clicking on candidate $y$ given context $c$ is modeled as a non-linear interaction function:

\begin{equation}
    P(\text{click} | u, y, c) = \sigma \left( \Phi_{\theta} \left( \mathbf{E}(\mathbf{x}_u) \parallel \mathbf{E}(\mathbf{x}_y) \parallel \mathcal{I}(\mathbf{x}_u, \mathbf{x}_y) \right) \right)
\end{equation}

where $\parallel$ represents vector concatenation, $\mathcal{I}(\cdot)$ denotes explicit feature interaction operations (e.g., cross products), $\Phi_{\theta}$ represents a parameterized Deep Neural Network (DNN) that learns high-order non-linear combinations, and $\sigma(\cdot)$ is the sigmoid function. 
Finally, the optimal headline $\hat{y}$ is selected by maximizing this predicted probability:
\begin{equation}
    \hat{y} = \argmax_{y \in \mathcal{Y}} P(\text{click} | u, y, c)
\end{equation}

By explicitly modeling features like education, this ranker captures personalized nuances—such as a specific segment preferring curiosity-driven titles over informative ones—that pure Bandit algorithms or simple semantic matching cannot resolve.

\section{Experiments and Results}


\subsection{Experimental Setup}


To evaluate robustness across heterogeneous scenarios, we curated MSR-50k, a benchmark dataset comprising 50,000 high-quality resources spanning graphics, short videos, and micro-videos across over 40 vertical categories (e.g., Drama, History). For online verification, models were deployed on a commercial platform with over 100 million DAUs.

\paragraph{Baselines.}

We adopt Qwen3-14B \cite{yang2025qwen3technicalreport} as the foundational backbone for our GESE framework. To rigorously evaluate the efficacy of our approach, we benchmark it against two distinct categories of baselines. Firstly, we employ massive-scale, state-of-the-art proprietary and open-weights models, including DeepSeek V3.1\cite{deepseekai2025deepseekv3technicalreport}, Qwen3-235B-A22B\cite{yang2025qwen3technicalreport}, and GPT-OSS-120B\cite{openai2025gptoss120bgptoss20bmodel}. Secondly, We include an SFT-Only baseline, fine-tuned exclusively on high-CTR online interaction logs. This represents the standard industrial practice of behavior cloning, serving as a direct reference to quantify the marginal gains achieved by our diversity-aware exploration mechanisms over simple supervised imitation.

\paragraph{Evaluation Metrics.}

To rigorously assess the trade-off between generative exploration and semantic precision, we employ a comprehensive evaluation protocol that scrutinizes the generated candidates from two complementary dimensions. 
We first quantify the breadth of the semantic space using Distinct-N ($N=1,2$)\cite{li2016diversitypromotingobjectivefunctionneural} to measure lexical uniqueness, while scrutinizing intra-list redundancy via Self-BLEU\cite{Zhu2018TexygenAB} and Pairwise-BLEU\cite{10.3115/1073083.1073135}; lower scores in these metrics indicate a more heterogeneous candidate set capable of covering diverse latent user interests. Concurrently, to ensure that high diversity does not compromise factual consistency, we evaluate content fidelity using ROUGE-1/2/L\cite{lin-2004-rouge} for surface-level alignment with editorial ground truth, and employ a robust NLI-based entailment metric \cite{yoran2024makingretrievalaugmentedlanguagemodels} to detect and penalize hallucinations, thereby guaranteeing that the generated headlines remain faithful to the source content.




\subsection{Online A/B Test}

We evaluated the framework on a information feed platform serving approximately 100 million Daily Active Users (DAU). Results are summarized in Table \ref{tab:online_results}.
The standard SFT-Only baseline underperformed the production baseline (-1.5\% CTR), confirming that greedy decoding leads to mode collapse and safe but unengaging headlines. Coupling the SFT generator with a bandit selector (\textsc{SFT \& UCB}) reversed the trend (+1.37\%), proving that selection is essential for filtering suboptimal candidates. Crucially, replacing the SFT generator with our GSPO explorer (\textsc{GSPO \& UCB}) yielded a further jump to +2.28\%, validating our core hypothesis that diversity is a prerequisite for effective personalization.

Finally, the full GESE framework, which integrates the diversity-aware generator with the real-time feedback deep selection model, achieved the highest performance across all metrics, yielding a 2.57\% increase in CTR and a 0.79\% increase in effective impressions. Crucially, this gain in click probability is accompanied by a 0.87\% increase in User Dwell Time. The simultaneous improvement in CTR and Dwell Time confirms that our hierarchical reward shaping successfully aligns the model with genuine user utility. It indicates that the generated headlines are not merely attractive "clickbait" but are faithfully aligned with the content, leading to sustained user engagement and satisfaction.

\begin{table}[h]
\centering
\caption{Online A/B testing results relative to Raw Headline. \textbf{GESE} achieves the best performance across all metrics.}
\label{tab:online_results}
\vspace{1mm}
\resizebox{\columnwidth}{!}{%
\begin{tabular}{lccc}
\toprule
\textbf{Methods} & \textbf{CTR Lift} & \textbf{Impressions} & \textbf{Dwell Time} \\
\midrule
Raw Headline & -- & -- & -- \\
SFT-Only & -1.50\% & N/A & N/A \\
SFT \& UCB & +1.37\% & +0.18\% & +0.33\% \\
GSPO \& UCB & +2.28\% & +0.53\% & +0.60\% \\
\textbf{GESE (Ours)} & \textbf{+2.57\%} & \textbf{+0.79\%} & \textbf{+0.87\%} \\
\bottomrule
\end{tabular}%
}
\end{table}

\subsection{Offline Evaluation}

\begin{table*}[t]
    \centering
    \caption{Offline performance comparison on the MSR-50k dataset. $\downarrow$ implies lower is better, while $\uparrow$ implies higher is better. \textbf{Bold} denotes the best performance within the Qwen3-14B group, and \underline{underline} denotes the best overall performance across all parameter scales.}
    \label{tab:offline_results}
    \resizebox{1.0\textwidth}{!}{
    \begin{tabular}{lcccccccc}
        \toprule
        \raisebox{-1.5ex}{\textbf{Model}} & \multicolumn{3}{c}{\textbf{Diversity Metrics}} & \multicolumn{4}{c}{\textbf{Quality \& Fidelity}} & \textbf{Appeal} \\
        \cmidrule(lr){2-4} \cmidrule(lr){5-8} \cmidrule(lr){9-9}
         & \textbf{\shortstack{Pairwise\\BLEU}} $\downarrow$ & \textbf{\shortstack{Self\\BLEU}} $\downarrow$ & \textbf{\shortstack{Distinct\\N-Gram}} $\uparrow$ & \textbf{\shortstack{NLI\\Score}} $\uparrow$ & \textbf{ROUGE-1} $\uparrow$ & \textbf{ROUGE-2} $\uparrow$ & \textbf{ROUGE-L} $\uparrow$ & \textbf{\shortstack{CTR\\Score}} $\uparrow$ \\
        \midrule
        \multicolumn{9}{l}{\textit{Base: Large Scale LLMs (Closed/Open)}} \\
        GPT-OSS-120B & \underline{3.92} & \underline{7.96} & 52.89 & 70.48 & 34.91 & 13.27 & 30.38 & 10.57 \\
        Qwen3-235B & 8.52 & 18.76 & 55.97 & 62.31 & \underline{50.24} & \underline{26.68} & \underline{44.77} & 7.21 \\
        DeepSeek-V3.1 & 4.63 & 12.80 & 51.37 & \underline{82.93} & 37.65 & 16.42 & 33.45 & 9.88 \\
        \midrule
        \multicolumn{9}{l}{\textit{Base: Qwen3-14B (Ours)}} \\
        SFT-Only & 8.98 & 18.02 & 49.40 & 75.66 & 48.04 & 26.34 & 44.09 & 9.16 \\
        \textbf{GESE (Ours)} & \textbf{4.47} & \textbf{9.93} & \textbf{\underline{56.20}} & \textbf{80.17} & 34.82 & 15.62 & 30.92 & \textbf{\underline{11.38}} \\
        \bottomrule
    \end{tabular}
    }
\end{table*}

\begin{table*}[h]
    \centering
    \caption{Ablation studies on reward functions and training stages. w/o Quality: Removes the faithfulness constraints. w/o Diversity: Removes the set-wise IoU reward. w/o RL: Corresponds to the SFT-Only baseline. w/o SFT: Applies GSPO directly to the pre-trained base model.}
    \label{tab:ablation}
    \resizebox{1.0\textwidth}{!}{
    \begin{tabular}{lcccccccc}
        \toprule
        \raisebox{-1.5ex}{\textbf{Method}} & \multicolumn{3}{c}{\textbf{Diversity Metrics}} & \multicolumn{4}{c}{\textbf{Quality \& Fidelity}} & \textbf{Appeal} \\
        \cmidrule(lr){2-4} \cmidrule(lr){5-8} \cmidrule(lr){9-9}
         & \textbf{\shortstack{Pairwise\\BLEU}} $\downarrow$ & \textbf{\shortstack{Self\\BLEU}} $\downarrow$ & \textbf{\shortstack{Distinct\\N-Gram}} $\uparrow$ & \textbf{\shortstack{NLI\\Score}} $\uparrow$ & \textbf{ROUGE-1} $\uparrow$ & \textbf{ROUGE-2} $\uparrow$ & \textbf{ROUGE-L} $\uparrow$ & \textbf{\shortstack{CTR\\Score}} $\uparrow$ \\
        \midrule
        \textbf{GESE (Ours)} & \textbf{4.47} & \textbf{9.93} & \textbf{56.20} & \textbf{80.17} & 34.82 & 15.62 & 30.92 & 11.38 \\
        \midrule
        \multicolumn{9}{l}{\textit{Ablation: Hierarchical Reward Functions}} \\
        \quad w/o Quality Constraint & 7.35 & 12.68 & 53.48 & 47.32 & 17.64 & 7.32 & 11.65 & \textbf{12.79} \\
        \quad w/o Diversity Bonus & 9.36 & 19.73 & 55.84 & 79.32 & 34.93 & 15.72 & 31.06 & 11.29 \\
        \quad w/o Both & 9.27 & 18.68 & 55.32 & 44.32 & 16.25 & 7.17 & 10.86 & \underline{12.86} \\
        \midrule
        \multicolumn{9}{l}{\textit{Ablation: Training Components}} \\
        \quad w/o RL (SFT-Only) & 8.98 & 18.02 & 49.40 & 75.66 & \textbf{48.04} & \textbf{26.34} & \textbf{44.09} & 9.16 \\
        \quad w/o SFT (Cold RL) & 4.51 & 10.62 & 37.83 & 74.39 & 35.49 & 14.37 & 29.85 & 10.19 \\
        \bottomrule
    \end{tabular}
    }
\end{table*}

Table \ref{tab:offline_results} presents the comparative performance across three critical dimensions: Diversity, Quality, and Appeal. As hypothesized, the SFT-Only baseline exhibits severe mode collapse, evidenced by high Pair-BLEU (8.98) and Self-BLEU (18.02) scores, comparable to the significantly larger Qwen3-235B. This confirms that supervised likelihood maximization tends to converge on repetitive patterns. In contrast, GESE reduces Self-BLEU by nearly 45\% (18.02 $\to$ 9.93) and achieves the highest Distinct-N-Gram score (56.20) among all models. Notably, our 14B model rivals the diversity metrics of the 120B-scale GPT model, validating that the GSPO mechanism effectively forces the explorer to traverse the semantic latent space rather than memorizing training modes. 

An intriguing observation is the divergence between ROUGE and NLI scores. SFT-Only achieves high ROUGE scores (e.g., R-L 44.09) but a lower NLI score (75.66). This indicates a "Parroting Effect," where the model mechanically copies the original title without understanding. conversely, GESE shows lower ROUGE scores yet maintains a high NLI (80.17). This trade-off is desirable: it demonstrates that GESE generates semantically novel headlines (low lexical overlap with ground truth) that remain factually consistent (high entailment) with the content, successfully avoiding the hallucination trap while providing fresh perspectives.

Most critically, GESE achieves the highest predicted CTR Score (11.38), outperforming both the SFT baseline (9.16) and even the 10x larger teacher models. This substantial margin underscores the efficacy of our hierarchical reward shaping, which aligns the generation directly with industrial engagement signals rather than mere linguistic plausibility.

\subsection{Ablation Study}

To disentangle the contribution of each component in the GESE framework, we conducted a series of ablation studies on the MSR-50k dataset. The results are summarized in Table \ref{tab:ablation}. A striking observation is that removing the Quality constraint (w/o Quality) results in the highest CTR Score (12.79), surpassing even our full model. However, this comes at a catastrophic cost: the NLI score plummets to 47.32, and ROUGE-L drops to 11.65. This phenomenon illustrates Reward Hacking: without semantic guardrails, the RL algorithm exploits the CTR model's biases, generating "clickbait" or hallucinations that differ wildly from the content but maximize predicted clicks. The Quality Model is thus proven essential not for engagement, but for maintaining the system's industrial viability and trustworthiness.

Ablating the diversity bonus (w/o Diversity) leads to a significant regression in exploration metrics, with Self-BLEU doubling from 9.93 to 19.73. This confirms that even with Group Sequence sampling, the model tends to collapse into a single high-reward mode without explicit set-wise penalties. Interestingly, the CTR score slightly decreases (11.29 vs. 11.38), suggesting that lack of diversity hinders the model from finding the global optimum in the reward landscape.

Comparing the component ablations reveals the necessity of the two-stage pipeline. w/o RL (SFT-Only) suffers from low appeal (CTR 9.16) and high redundancy. Conversely, w/o SFT (Cold RL) exhibits degradation in linguistic quality, evidenced by the lowest Distinct-N-Gram (37.83) and ROUGE scores. This indicates that SFT is crucial for establishing the structural grammar of the task, while RL is required to align this capability with user preference.

\section{Conclusion}

In this work, we challenge the conventional point-estimation paradigm in personalized generation, demonstrating that forcing LLMs to converge on a single optimal output under profile uncertainty inevitably leads to mode collapse. We introduce GESE, a decoupled framework that redefines the task as generative exploration via Group Sequence Policy Optimization (GSPO) followed by selective exploitation. By optimizing for the coverage of latent interest modes rather than mere token likelihood, our approach effectively functions as a mathematical hedge against the aleatoric uncertainty inherent in industrial user profiles. Extensive deployment on a commercial platform with 100 million DAUs confirms that this paradigm significantly boosts both engagement (CTR) and user satisfaction (Dwell Time), providing a generalized blueprint for robustly aligning generative AI with dynamic user utility in stochastic environments.

\bibliographystyle{named}
\bibliography{references}

@inproceedings{10.1145/3626772.3657684,
author = {Zhang, Kaike and Cao, Qi and Wu, Yunfan and Sun, Fei and Shen, Huawei and Cheng, Xueqi},
title = {LoRec: Combating Poisons with Large Language Model for Robust Sequential Recommendation},
year = {2024},
isbn = {9798400704314},
publisher = {Association for Computing Machinery},
address = {New York, NY, USA},
url = {https://doi.org/10.1145/3626772.3657684},
doi = {10.1145/3626772.3657684},
booktitle = {Proceedings of the 47th International ACM SIGIR Conference on Research and Development in Information Retrieval},
pages = {1733–1742},
numpages = {10},
location = {Washington DC, USA},
series = {SIGIR '24}
}

@inproceedings{jiang-etal-2025-reclm,
    title = "{R}ec{LM}: Recommendation Instruction Tuning",
    author = "Jiang, Yangqin  and
      Yang, Yuhao  and
      Xia, Lianghao  and
      Luo, Da  and
      Lin, Kangyi  and
      Huang, Chao",
    editor = "Che, Wanxiang  and
      Nabende, Joyce  and
      Shutova, Ekaterina  and
      Pilehvar, Mohammad Taher",
    booktitle = "Proceedings of the 63rd Annual Meeting of the Association for Computational Linguistics (Volume 1: Long Papers)",
    month = jul,
    year = "2025",
    address = "Vienna, Austria",
    publisher = "Association for Computational Linguistics",
    url = "https://aclanthology.org/2025.acl-long.751/",
    doi = "10.18653/v1/2025.acl-long.751",
    pages = "15443--15459",
    ISBN = "979-8-89176-251-0"
}

@inproceedings{Li2024PreservingDI,
  title={Preserving Diversity in Supervised Fine-Tuning of Large Language Models},
  author={Ziniu Li and Congliang Chen and Tian Xu and Zeyu Qin and Jiancong Xiao and Ruoyu Sun and Zhimin Luo},
  booktitle={International Conference on Learning Representations},
  year={2024},
  url={https://api.semanticscholar.org/CorpusID:272146685}
}

@inproceedings{10.5555/3666122.3668460,
author = {Rafailov, Rafael and Sharma, Archit and Mitchell, Eric and Ermon, Stefano and Manning, Christopher D. and Finn, Chelsea},
title = {Direct preference optimization: your language model is secretly a reward model},
year = {2023},
publisher = {Curran Associates Inc.},
address = {Red Hook, NY, USA},
booktitle = {Proceedings of the 37th International Conference on Neural Information Processing Systems},
articleno = {2338},
numpages = {14},
location = {New Orleans, LA, USA},
series = {NIPS '23}
}

@inproceedings{NEURIPS2024_f25d75fc,
 author = {Miao, Yuchun and Zhang, Sen and Ding, Liang and Bao, Rong and Zhang, Lefei and Tao, Dacheng},
 booktitle = {Advances in Neural Information Processing Systems},
 doi = {10.52202/079017-4270},
 editor = {A. Globerson and L. Mackey and D. Belgrave and A. Fan and U. Paquet and J. Tomczak and C. Zhang},
 pages = {134387--134429},
 publisher = {Curran Associates, Inc.},
 title = {InfoRM: Mitigating Reward Hacking in RLHF via Information-Theoretic Reward Modeling},
 url = {https://proceedings.neurips.cc/paper_files/paper/2024/file/f25d75fc760aec0a6174f9f5d9da59b8-Paper-Conference.pdf},
 volume = {37},
 year = {2024}
}

@misc{zheng2025groupsequencepolicyoptimization,
      title={Group Sequence Policy Optimization}, 
      author={Chujie Zheng and Shixuan Liu and Mingze Li and Xiong-Hui Chen and Bowen Yu and Chang Gao and Kai Dang and Yuqiong Liu and Rui Men and An Yang and Jingren Zhou and Junyang Lin},
      year={2025},
      eprint={2507.18071},
      archivePrefix={arXiv},
      primaryClass={cs.LG},
      url={https://arxiv.org/abs/2507.18071}, 
}

@ARTICLE{10506571,
  author={Zhao, Zihuai and Fan, Wenqi and Li, Jiatong and Liu, Yunqing and Mei, Xiaowei and Wang, Yiqi and Wen, Zhen and Wang, Fei and Zhao, Xiangyu and Tang, Jiliang and Li, Qing},
  journal={IEEE Transactions on Knowledge and Data Engineering}, 
  title={Recommender Systems in the Era of Large Language Models (LLMs)}, 
  year={2024},
  volume={36},
  number={11},
  pages={6889-6907},
  doi={10.1109/TKDE.2024.3392335}}

@misc{geng2023recommendationlanguageprocessingrlp,
      title={Recommendation as Language Processing (RLP): A Unified Pretrain, Personalized Prompt \& Predict Paradigm (P5)}, 
      author={Shijie Geng and Shuchang Liu and Zuohui Fu and Yingqiang Ge and Yongfeng Zhang},
      year={2023},
      eprint={2203.13366},
      archivePrefix={arXiv},
      primaryClass={cs.IR},
      url={https://arxiv.org/abs/2203.13366}, 
}

@InProceedings{pmlr-v235-ethayarajh24a,
  title = 	 {Model Alignment as Prospect Theoretic Optimization},
  author =       {Ethayarajh, Kawin and Xu, Winnie and Muennighoff, Niklas and Jurafsky, Dan and Kiela, Douwe},
  booktitle = 	 {Proceedings of the 41st International Conference on Machine Learning},
  pages = 	 {12634--12651},
  year = 	 {2024},
  editor = 	 {Salakhutdinov, Ruslan and Kolter, Zico and Heller, Katherine and Weller, Adrian and Oliver, Nuria and Scarlett, Jonathan and Berkenkamp, Felix},
  volume = 	 {235},
  series = 	 {Proceedings of Machine Learning Research},
  month = 	 {21--27 Jul},
  publisher =    {PMLR},
  url = 	 {https://proceedings.mlr.press/v235/ethayarajh24a.html}
}

@inproceedings{zhou-etal-2024-beyond,
    title = "Beyond One-Preference-Fits-All Alignment: Multi-Objective Direct Preference Optimization",
    author = "Zhou, Zhanhui  and
      Liu, Jie  and
      Shao, Jing  and
      Yue, Xiangyu  and
      Yang, Chao  and
      Ouyang, Wanli  and
      Qiao, Yu",
    editor = "Ku, Lun-Wei  and
      Martins, Andre  and
      Srikumar, Vivek",
    booktitle = "Findings of the Association for Computational Linguistics: ACL 2024",
    month = aug,
    year = "2024",
    address = "Bangkok, Thailand",
    publisher = "Association for Computational Linguistics",
    url = "https://aclanthology.org/2024.findings-acl.630/",
    doi = "10.18653/v1/2024.findings-acl.630",
    pages = "10586--10613"
}

@inproceedings{Wang_2025, 
   series={WWW ’25},
   title={Unleashing the Power of Large Language Model for Denoising Recommendation},
   url={http://dx.doi.org/10.1145/3696410.3714758},
   DOI={10.1145/3696410.3714758},
   booktitle={Proceedings of the ACM on Web Conference 2025},
   publisher={ACM},
   author={Wang, Shuyao and Zheng, Zhi and Sui, Yongduo and Xiong, Hui},
   year={2025},
   month=apr, pages={252–263},
   collection={WWW ’25} }

@article{articleATI,
author = {Zhao, Yuying and Wang, Yu and Liu, Yunchao and Cheng, Xueqi and Aggarwal, Charu and Derr, Tyler},
year = {2024},
month = {05},
pages = {},
title = {Fairness and Diversity in Recommender Systems: A Survey},
volume = {16},
journal = {ACM Transactions on Intelligent Systems and Technology},
doi = {10.1145/3664928}
}

@inproceedings{10.5555/3327345.3327465,
author = {Chen, Laming and Zhang, Guoxin and Zhou, Hanning},
title = {Fast greedy MAP inference for determinantal point process to improve recommendation diversity},
year = {2018},
publisher = {Curran Associates Inc.},
address = {Red Hook, NY, USA},
booktitle = {Proceedings of the 32nd International Conference on Neural Information Processing Systems},
pages = {5627–5638},
numpages = {12},
location = {Montr\'{e}al, Canada},
series = {NIPS'18}
}

@misc{zhang2025moslimaligndiversepreferencesprompts,
      title={MOSLIM:Align with diverse preferences in prompts through reward classification}, 
      author={Yu Zhang and Wanli Jiang and Zhengyu Yang},
      year={2025},
      eprint={2505.20336},
      archivePrefix={arXiv},
      primaryClass={cs.CL},
      url={https://arxiv.org/abs/2505.20336}, 
}

@misc{ye2024scalarrewardmodellearning,
      title={Beyond Scalar Reward Model: Learning Generative Judge from Preference Data}, 
      author={Ziyi Ye and Xiangsheng Li and Qiuchi Li and Qingyao Ai and Yujia Zhou and Wei Shen and Dong Yan and Yiqun Liu},
      year={2024},
      eprint={2410.03742},
      archivePrefix={arXiv},
      primaryClass={cs.CL},
      url={https://arxiv.org/abs/2410.03742}, 
}

@inproceedings{10.5555/3692070.3694392,
author = {Yang, Rui and Pan, Xiaoman and Luo, Feng and Qiu, Shuang and Zhong, Han and Yu, Dong and Chen, Jianshu},
title = {Rewards-in-context: multi-objective alignment of foundation models with dynamic preference adjustment},
year = {2024},
publisher = {JMLR.org},
booktitle = {Proceedings of the 41st International Conference on Machine Learning},
articleno = {2322},
numpages = {22},
location = {Vienna, Austria},
series = {ICML'24}
}

@inproceedings{zhou-etal-2025-balancing,
    title = "Balancing Diversity and Risk in {LLM} Sampling: How to Select Your Method and Parameter for Open-Ended Text Generation",
    author = "Zhou, Yuxuan  and
      Keuper, Margret  and
      Fritz, Mario",
    editor = "Che, Wanxiang  and
      Nabende, Joyce  and
      Shutova, Ekaterina  and
      Pilehvar, Mohammad Taher",
    booktitle = "Proceedings of the 63rd Annual Meeting of the Association for Computational Linguistics (Volume 1: Long Papers)",
    month = jul,
    year = "2025",
    address = "Vienna, Austria",
    publisher = "Association for Computational Linguistics",
    url = "https://aclanthology.org/2025.acl-long.1278/",
    doi = "10.18653/v1/2025.acl-long.1278",
    pages = "26352--26365",
    ISBN = "979-8-89176-251-0"
}

@inproceedings{10.5555/3737916.3741662,
author = {Tang, Hao and Hu, Keya and Zhou, Jin Peng and Zhong, Sicheng and Zheng, Wei-Long and Si, Xujie and Ellis, Kevin},
title = {Code repair with LLMs gives an exploration-exploitation tradeoff},
year = {2024},
isbn = {9798331314385},
publisher = {Curran Associates Inc.},
address = {Red Hook, NY, USA},
booktitle = {Proceedings of the 38th International Conference on Neural Information Processing Systems},
articleno = {3746},
numpages = {43},
location = {Vancouver, BC, Canada},
series = {NIPS '24}
}

@misc{yang2025qwen3technicalreport,
      title={Qwen3 Technical Report}, 
      author={Qwen},
      year={2025},
      eprint={2505.09388},
      archivePrefix={arXiv},
      primaryClass={cs.CL},
      url={https://arxiv.org/abs/2505.09388}, 
}

@misc{deepseekai2025deepseekv3technicalreport,
      title={DeepSeek-V3 Technical Report}, 
      author={DeepSeek-AI},
      year={2025},
      eprint={2412.19437},
      archivePrefix={arXiv},
      primaryClass={cs.CL},
      url={https://arxiv.org/abs/2412.19437}, 
}

@misc{openai2025gptoss120bgptoss20bmodel,
      title={gpt-oss-120b \& gpt-oss-20b Model Card}, 
      author={OpenAI},
      year={2025},
      eprint={2508.10925},
      archivePrefix={arXiv},
      primaryClass={cs.CL},
      url={https://arxiv.org/abs/2508.10925}, 
}

@misc{li2016diversitypromotingobjectivefunctionneural,
      title={A Diversity-Promoting Objective Function for Neural Conversation Models}, 
      author={Jiwei Li and Michel Galley and Chris Brockett and Jianfeng Gao and Bill Dolan},
      year={2016},
      eprint={1510.03055},
      archivePrefix={arXiv},
      primaryClass={cs.CL},
      url={https://arxiv.org/abs/1510.03055}, 
}

@article{Zhu2018TexygenAB,
  title={Texygen: A Benchmarking Platform for Text Generation Models},
  author={Yaoming Zhu and Sidi Lu and Lei Zheng and Jiaxian Guo and Weinan Zhang and Jun Wang and Yong Yu},
  journal={The 41st International ACM SIGIR Conference on Research \& Development in Information Retrieval},
  year={2018},
  url={https://api.semanticscholar.org/CorpusID:3636178}
}

@inproceedings{10.3115/1073083.1073135,
author = {Papineni, Kishore and Roukos, Salim and Ward, Todd and Zhu, Wei-Jing},
title = {BLEU: a method for automatic evaluation of machine translation},
year = {2002},
publisher = {Association for Computational Linguistics},
address = {USA},
url = {https://doi.org/10.3115/1073083.1073135},
doi = {10.3115/1073083.1073135},
booktitle = {Proceedings of the 40th Annual Meeting on Association for Computational Linguistics},
pages = {311–318},
numpages = {8},
location = {Philadelphia, Pennsylvania},
series = {ACL '02}
}

@misc{yoran2024makingretrievalaugmentedlanguagemodels,
      title={Making Retrieval-Augmented Language Models Robust to Irrelevant Context}, 
      author={Ori Yoran and Tomer Wolfson and Ori Ram and Jonathan Berant},
      year={2024},
      eprint={2310.01558},
      archivePrefix={arXiv},
      primaryClass={cs.CL},
      url={https://arxiv.org/abs/2310.01558}, 
}

@inproceedings{lin-2004-rouge,
    title = "{ROUGE}: A Package for Automatic Evaluation of Summaries",
    author = "Lin, Chin-Yew",
    booktitle = "Text Summarization Branches Out",
    month = jul,
    year = "2004",
    address = "Barcelona, Spain",
    publisher = "Association for Computational Linguistics",
    url = "https://aclanthology.org/W04-1013/",
    pages = "74--81"
}

\end{document}